\documentclass[prl,aps,twocolumn,showpacs,superscriptaddress,floatfix,amsmath,amssymb,nofootinbib]{revtex4}
\usepackage{amsfonts}
\usepackage{graphicx}
\newcommand{\beq}{\begin{equation}}
\newcommand{\eeq}{\end{equation}}

\newcommand{\beqa}{\begin{eqnarray}}
\newcommand{\eeqa}{\end{eqnarray}}
\newcommand{\beqax}{\begin{eqnarray*}}
\newcommand{\eeqax}{\end{eqnarray*}}

\def\ra{\rangle}
\def\la{\langle}
\begin{document}
\title{Engineering fast and stable splitting of matter waves}
%
\author{E. Torrontegui}
\affiliation{Departamento de Qu\'{\i}mica F\'{\i}sica, Universidad del Pa\'{\i}s Vasco - Euskal Herriko Unibertsitatea, 
Apdo. 644, Bilbao, Spain}

\author{S. Mart\'{\i}nez-Garaot}
\affiliation{Departamento de Qu\'{\i}mica F\'{\i}sica, Universidad del Pa\'{\i}s Vasco - Euskal Herriko Unibertsitatea, 
Apdo. 644, Bilbao, Spain}

\author{M. Modugno}
\affiliation{Departamento de F\'{\i}sica Te\'orica e Historia de la Ciencia, Universidad del Pa\'{\i}s Vasco - Euskal Herriko Unibertsitatea, 
Apdo. 644, Bilbao, Spain}
\affiliation{IKERBASQUE, Basque Foundation for Science, Alameda Urquijo 36, 48011 Bilbao, Spain}

\author{Xi Chen} 
\affiliation{Departamento de Qu\'{\i}mica F\'{\i}sica, Universidad del Pa\'{\i}s Vasco - Euskal Herriko Unibertsitatea, 
Apdo. 644, Bilbao, Spain}
\affiliation{Department of Physics, Shanghai University, 200444 Shanghai, People's Republic of China}

\author{J. G. Muga}
\affiliation{Departamento de Qu\'{\i}mica F\'{\i}sica, Universidad del Pa\'{\i}s Vasco - Euskal Herriko Unibertsitatea, 
Apdo. 644, Bilbao, Spain}
\affiliation{Department of Physics, Shanghai University, 200444 Shanghai, People's Republic of China}
\begin{abstract}
When attempting to split coherent cold atom clouds or a Bose-Einstein condensate (BEC) by bifurcation of the trap into a double well, slow adiabatic following is unstable with respect to any slight asymmetry, and the wave ``collapses'' to the lower well, 
whereas a generic fast chopping splits the wave but it also excites it.
Shortcuts to adiabaticity engineered to speed up the adiabatic process through non-adiabatic transients, provide instead quiet and robust fast splitting.  
The non-linearity of the BEC makes the proposed shortcut even more stable. 
\end{abstract}  	
\maketitle
%
%
%
%
%
%
{\it Introduction.---} The splitting of a wavefunction is an important operation for  
matter wave interferometry \cite{S07,S09a,S09b,Augusto}.
It is a peculiar one though, as adiabatic following, rather than being robust, is intrinsically unstable with respect to a small external potential asymmetry \cite{JGB}. The ground-state wavefunction ``collapses'' into the slightly lower well so that a very slow trap potential bifurcation in fact fails to split the wave except 
for perfectly symmetrical potentials.  
An arbitrarily fast bifurcation may remedy this but at the price of
a strong excitation which is also undesired. 
We propose here a way out to these problems
by using shortcuts to adiabaticity that speed up the adiabatic process along a non-adiabatic route. The wave splitting via shortcuts avoids the final excitation 
and turns out to be signifficantly more stable than the adiabatic following with respect to the asymmetric perturbation.  
Specifically we shall use a simple inversion method: a streamlined version \cite{ErikFF} of the fast-forward technique of Masuda and Nakamura \cite{MNProc} applied to Gross-Pitaievski (GP) or Schr\"odinger equations.   
We have previously found some 
obstacles to apply the invariants-based method (at least using quadratic-in momentum invariants \cite{ErikFF}) and the transitionless-driving algorythm \cite{Rice} (because of difficulties to implement in practice the counter-diabatic terms).

{\it{Fast-forward approach.}---}
The fast-forward method \cite{MNPRA08,MNProc,ErikFF} 
may be used to generate external potentials to drive the matter wave from the initial single well to a final symmetric double well.
The starting point of the streamlined version in \cite{ErikFF} 
is the $3D$ time-dependent GP equation
\beq
\label{start}
i\hbar\frac{\partial|\psi(t)\rangle}{\partial t}=H(t)|\psi(t)\rangle,
\eeq
where the Hamiltonian $H(t)=T+G(t)+V(t)$ includes the  kinetic energy $T$, the external potential $V$,
and  the mean field potential $G$. Assuming that $V$ is local, $\langle \bold{x}|V(t)|\bold{x'}\rangle=V(\bold{x},t)\delta(\bold{x}-\bold{x'})$, it 
may be written from Eq. (\ref{start}) as 
\beq
\label{pot1}
V(\bold{x},t)=\frac{i\hbar\langle \bold{x}| \partial_t\psi(t)\rangle-\langle\bold{x}|T|\psi(t)\rangle-\langle\bold{x}|G(t)|\psi(t)\rangle}{\langle\bold{x}|\psi(t)\rangle}, 
\eeq 
with $\langle\bold{x}|\psi(t)\rangle=\psi(\bold{x},t)$, whereas
\beqax
\langle\bold{x}|T|\psi(t)\rangle&=&\frac{-\hbar^2}{2m} \nabla^2\psi(\bold{x},t),
\\
\langle\bold{x}|G(t)|\psi(t)\rangle&=&gN|\psi(\bold{x},t)|^2\psi(\bold{x},t), 
\eeqax
%
where $g$ is the coupling constant of the BEC and $N$ is the number of atoms. For the
numerical examples  we consider $^{87}$Rb atoms, $m=1.44\times 10^{-25}$ kg.
Using in Eq. (\ref{pot1}) the ansatz
\beq
\label{wave}
\langle\bold{x}|\psi(t)\rangle=r(\bold{x},t)e^{i\phi(\bold{x},t)}, \quad r(\bold{x},t), \phi(\bold{x},t) \in \mathbb{R},
\eeq
the real and imaginary parts of $V$ are
\beqa
{\rm{Re}}[V(\bold{x},t)]&=&-\hbar{\dot \phi}+\frac{\hbar^2}{2m}\bigg(\frac{\nabla^2 r}{r}-(\nabla \phi)^2\bigg)-gNr^2, \label{real}
\\
{\rm{Im}}[V(\bold{x},t)]&=&\hbar\frac{\dot r}{r}+\frac{\hbar^2}{2m}\bigg(\frac{2\nabla \phi\cdot \nabla r}{r}+\nabla^2 \phi\bigg),
\label{imag}
\eeqa
where the dot means time derivative.
We shall impose ${\rm{Im}}[V(\bold{x},t)]=0$
to design  a real potential. In addition 
we shall require that the ground state of the initial Hamiltonian $H(0)$ evolves in a time $t_f$ into the corresponding ground state of the final
$H(t_f)$, assuming that the Hamiltonian is known at the boundary times. 
  
In the inversion protocol, $r(\bold{x},t)$ is designed first,
and we solve for $\phi$ in Eq. (\ref{imag}) 
to get
$V_{FF}:={\rm{Re}}[V(\bold{x},t)]$ from Eq. (\ref{real}).  
To ensure that the initial and final states are eigenstates of the stationary GP equation we impose  
$\dot{r}=0$ at $t=0$ and $t_f$. Then Eq. (\ref{imag}) has
solutions $\phi(\bold{x},t)$ independent of $\bold{x}$ at the boundary times \cite{ErikFF}.  
Using this in Eq. (\ref{real}) at $t=0$, and multiplying by $e^{i\phi(0)}$, we get  
\beq
\bigg[-\frac{\hbar^2}{2m}\nabla^2+V(\bold x,0)+g|\psi(\bold x,0)|^2\bigg]\psi(\bold x,0)=-\hbar\dot \phi(0)\psi(\bold x,0). 
\eeq 
The initial state $\psi(\bold x,0)$ is an eigenstate of the stationary GP equation with chemical potential $-\hbar\dot\phi(0)=\mu(0)$.
A similar result is found at $t_f$.  

\begin{figure}[t]
\includegraphics[width=8.5cm]{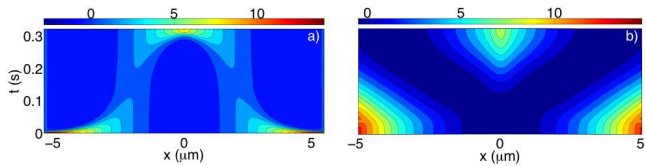}
\caption{(Color online) Contour plot of the fast forward potential $V_{FF}$ in units $\hbar\omega$ from Eq. (\ref{real}) for the interpolations made by (a) Eq. (\ref{ansatzr}),
and
(b) Eq. (\ref{ansatzrbueno}). Both interpolations produce the same initial and final states but {\itshape b)} produces a simpler $Y$-shape. Parameter
values: $\omega=780$ rad/s, $a=4$ $\mu$m and, $t_f=320$ ms.}
\label{f0}
\end{figure}

To illustrate this method we consider first a $1D$ linear Schr\"odinger equation ($g=0$) and apply the fast-forward approach to  split an initial single Gaussian 
state $r(x,0)=e^{-\Gamma^2 x^2/2}$ $(\Gamma=\sqrt{m\omega/\hbar})$ 
into a final double Gaussian $r(x,t_f)=e^{-\Gamma^2 (x-a)^2/2}+e^{-\Gamma^2(x+a)^2/2}$. 
In previous works \cite{MNProc,ErikFF} use has been made of the interpolation
\beq
\label{ansatzr}
r(x,t)=z(t)\bigg\{[1-{\cal{R}}(t)]r(x,0)+ 
{\cal{R}}(t)r(x,t_f)\bigg\},
\eeq
where ${\cal{R}}(t)$ is some smooth, monotonously increasing function
from 0 to 1 obeying $\dot {\cal R}=0$ so that $\dot r=0$ at the boundary times $t=0$ and $t_f$, and $z(t)$ is a normalization function.
This produces three wave-function bumps at intermediate times and a corresponding three-well potential. Here we use instead the two-bump form 
\beq
\label{ansatzrbueno}
r(x,t)=z(t)[e^{-\Gamma^2 (x-x_0(t))^2/2}+e^{-\Gamma^2 (x+x_0(t))^2/2}],
\eeq
which generates simpler $Y$-shaped potentials, see Fig. \ref{f0}. 
We also impose that $\dot{x}_0(0)=\dot{x}_0(t_f)=0$ so $\dot r=0$ at the boundary times.
In the numerical examples we impose for the Gaussian trajectory the polynomial $x_0(s)=a(3s^2-2 s^3)$, where $s=t/t_f$,  
and solve Eq. (\ref{imag}) with the initial conditions $\phi(x=0)=\frac{\partial{\phi}}{\partial x}|_{x=0}=0$ that fix the zero energy point.

%
%
%
%
%
%
%
%
{\it{Effect of the perturbation}.---}
\label{perturbation}
Now let us assume that a small asymmetry affects the splitting process. 
We model this with a potential $V_{\lambda}=V_{FF}+\lambda\theta(x)$,
where $\theta$ is the step function. 
The splitting becomes unstable, as we shall see, but the instability does not 
depend strongly on this particular form, which is chosen for simplicity. 
It would also be found for a linear-in-$x$ 
perturbation, a smoothed step, slightly different frequencies for the final right and left traps, or a displacement of the central barrier \cite{JGB}.    
%
%
%

\begin{figure}[t]
\includegraphics[width=6.cm]{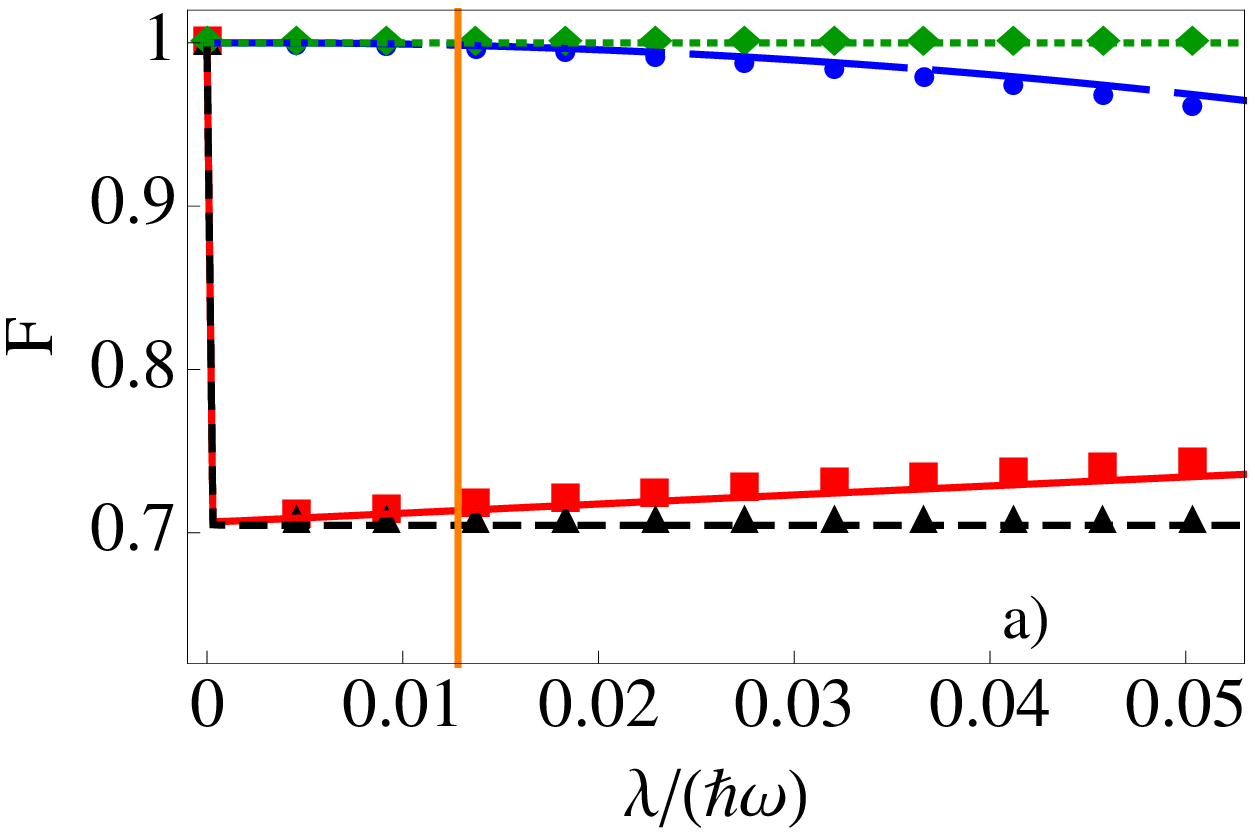}
\includegraphics[width=6.cm]{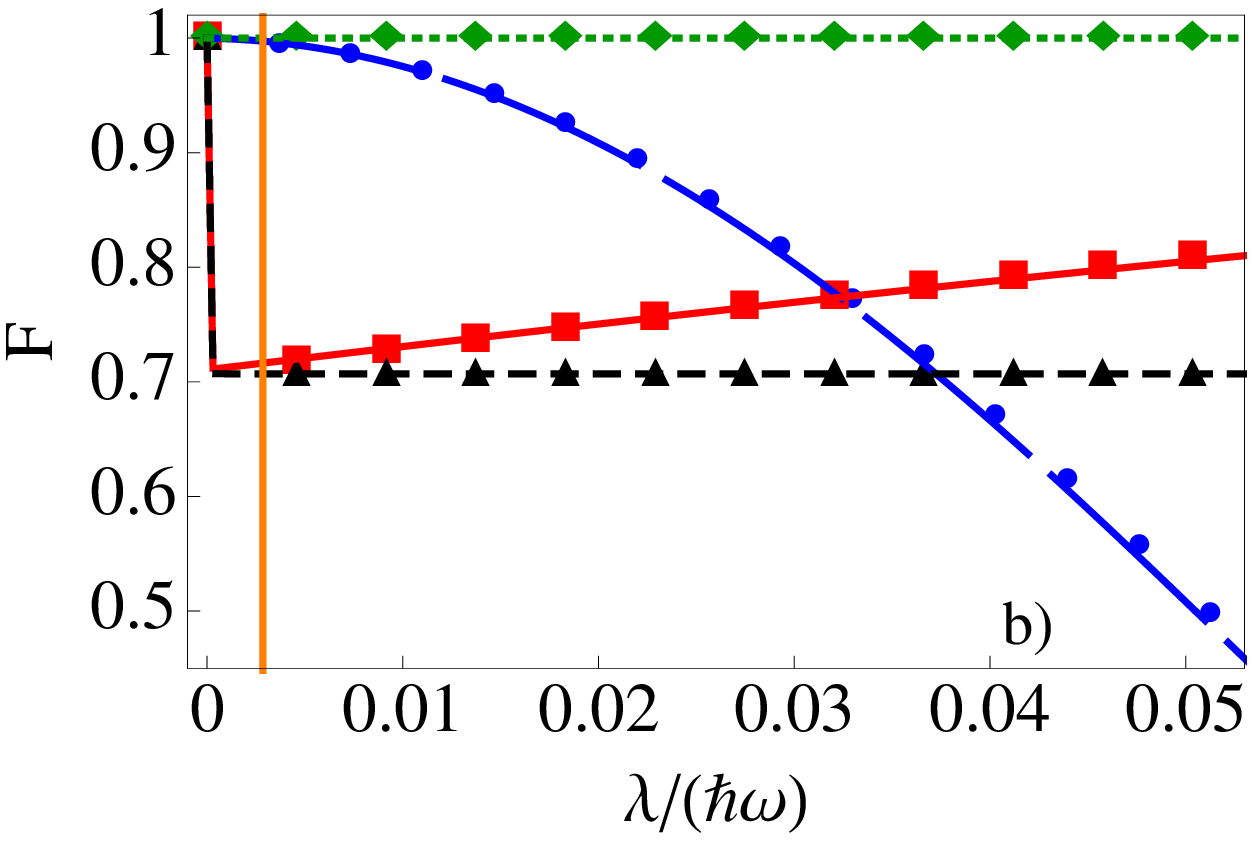}
\includegraphics[width=6.cm]{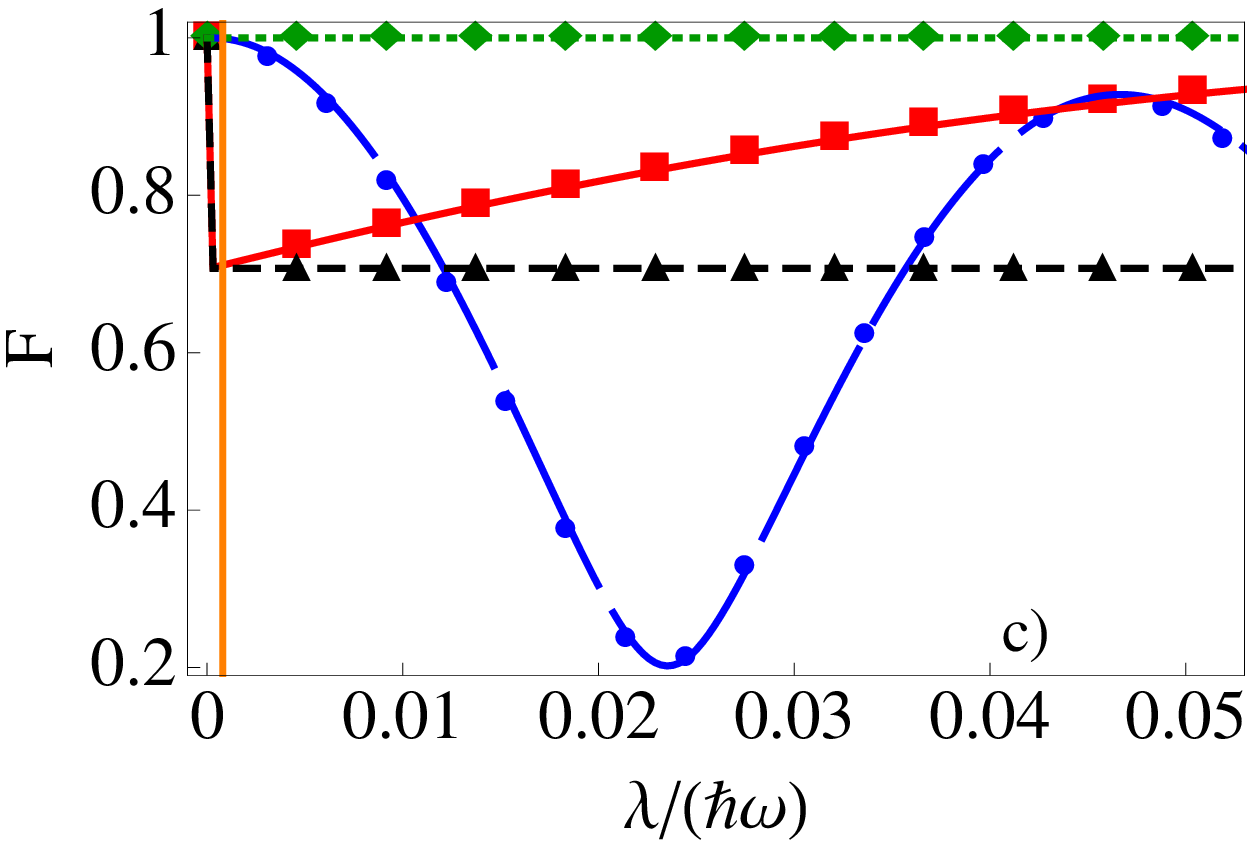}
\caption{(Color online) Different fidelities versus the perturbation parameter $\lambda$ for the fast-forward approach (lines) and the two-mode model (symbols). 
$F_D^{(0)}$: (blue) long-dashed line and circles; $F_D$: (red) solid line and squares;  $F_S$: (black) short-dashed line and triangles; $F_I$: (green) dotted line and rombs.
The (yellow) vertical line is at $0.2/t_f$. 
{(a)}  $t_f=20$ ms. {(b)} $t_f=90$ ms. {(c)} $t_f=320$ ms. The other  parameters are the same as in Fig. \ref{f0}.}
\label{f2}
\end{figure}

%
%
%
%
To analyze the effects of the perturbation $\lambda$ we compute several ``fidelities'': 
The black short-dashed line of Fig. \ref{f2} represents the structural fidelity $F_S=|\langle \psi^{-}_0(t_f)|\psi^{-}_\lambda(t_f)\rangle|$. 
It is the modulus of the overlap between the (perfectly split) ground state $\psi^{-}_0(t_f)$ of the unperturbed potential $V_{FF}(t_f)$ and the final ground state $ \psi^{-}_\lambda(t_f)$ of the 
actual, perturbed potential $V_{\lambda}$. This would be the fidelity found with the desired split state if the process were adiabatic. 
$F_S(\lambda)$ decays extremely rapidly from 1
at $\lambda=0$ to $1/\sqrt{2}$, which corresponds to the collapse 
of the ground state of the perturbed potential $V_{\lambda}$ into the deeper well. 

$F_D^{(0)}=|\langle \psi^-_0(t_f)|\psi(t_f)\rangle|$, the blue long-dashed line in Fig. \ref{f2},
is the modulus of 
the overlap between the state dynamically evolved with the perturbed potential $V_{\lambda}$,
$\psi(x,t_f)=\langle x|e^{iH_{\lambda}t_f/\hbar}|\psi(0)\rangle$, and $\psi^-_0(t_f)$, the final ground state of the unperturbed potential $V_{FF}(t_f)$. 
$\psi(0)=\psi^-_\lambda(0)$ is the
initial ground state with $V_{\lambda}(0)$, but the difference 
with using instead $\psi(0)=\psi^-_0(0)$ in the examples shown 
is negligible, as shown by the overlap $F_I=|\la \psi^-_\lambda(0)|\psi^-_0(0)\ra|\approx 1$, see the green dotted line in Fig. 2. 

 
%
%
%
%
%
%
%
%
The flatness of $F_D^{(0)}(\lambda)$ at small $\lambda$ is in sharp contrast to the rapid decay of $F_S(\lambda)$. 
In practice this feature enables us to perform robustly the desired splitting. Note that shorter process times $t_f$ make the splitting more stable, compare the Figs. \ref{f2}(a), (b), and (c).      

Finally, we also calculate $F_D=|\langle \psi(t_f)|\psi^{-}_\lambda(t_f)\rangle|$, the fidelity between the evolved state $\psi(t_f)$ and the final ground state $\psi^{-}_\lambda(t_f)$
for the perturbed potential $V_{\lambda}$ (red dotted line of Fig. \ref{f2}). 
For very small perturbations, $F_D\approx F_S$. In this regime the dynamical wave function $\psi(t_f)$ is not affected by the perturbation and becomes
$\psi^-_0(t_f)$, up to a phase factor, 
as confirmed also by the fact that $F_D^{(0)}\approx 1$ there. 
We shall understand and quantify this important regime below as a sudden process 
in a moving-frame interaction picture.  
As the perturbation $\lambda$ increases, the energy levels of 
the ground and excited states of $V_{\lambda}$ separate and the process 
becomes progressively less sudden and more adiabatic. In Fig. \ref{f2}(c) for  $t_f=320$ ms and for large values of $\lambda$, $F_D$ approaches 1 again, the final evolved state collapses to one side, and becomes the ground state of $V_{\lambda}$. 
For the shorter final times in Fig. \ref{f2}(a) and (b), 
larger $\lambda$ are needed to make $F_D$ approach 1 adiabatically. 
%
%
%
%
%

{\it{Moving two-mode model}.---}  
%
%
%
%
%
\begin{figure}
\includegraphics[height=4.cm]{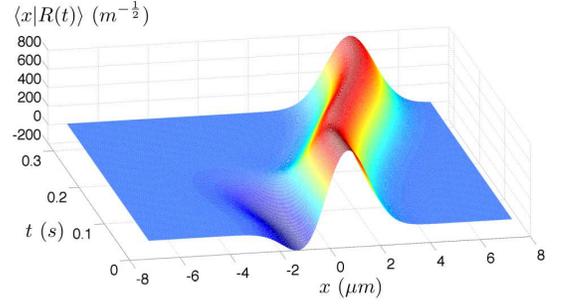}
\caption{(Color online)
Coordinate representation of the time dependent bare basis. The right vector $|R(t)\rangle$ is plotted for the parameters of Fig. \ref{f0}. The left vector $|L(t)\rangle$ satisfies $\la x|R(t)\ra=\la -x|L(t)\ra$ and is orthogonal to it. The side (negative) peak eventually dissapears.  
}
\label{f3}
\end{figure}
%
%
%
%
Static two-mode models have been previously used 
to analyze splitting processes \cite{Javanainen99,Schmiedmayer09,Aichmayr}. 
Here we add the separation motion of left and right basis functions to 
provide
analytical estimates and insight.  
In terms of a (moving) orthogonal bare basis
$|L(t)\rangle = \left( \begin{array} {rccl} 0\\ 1 \end{array} \right)$, $|R(t)\rangle = \left( \begin{array} {rccl} 1\\ 0 \end{array} \right)$ 
our two-mode Hamiltonian model is
\beq
\label{H_tm}
H(t)=\frac{\hbar}{2} \left ( \begin{array}{cc}
\lambda
& -\delta(t)\\
-\delta(t)& -\lambda
\end{array} \right),
\eeq 
where $\delta(t)$ is the tunneling rate \cite{Javanainen99, Schmiedmayer09} and $\lambda$ the  energy difference between the depths of the two wells \cite{Aichmayr}.
We may simply consider $\lambda$ constant through a given splitting process for the time being, and equal to the perturbative parameter that defines the asymmetry.  
A more detailed approach that we shall describe later will not produce any significant difference.        
The instantaneous eigenvalues are
\beq
\label{eigenvalues_tm}
E^{\pm}_\lambda(t)=
\pm \frac{\hbar}{2} \sqrt{\lambda^2+\delta^2(t)},
\eeq
and the normalized eigenstates 
\beq
\begin{array}{ll}
\label{eigenstates_tm}
|\psi^+_\lambda(t)\rangle = \sin{ \left ( \frac{\alpha}{2} \right ) } |L(t)\rangle-\cos {\left ( \frac{\alpha}{2} \right ) }|R(t)\rangle, 
\\
\\
|\psi^-_\lambda(t)\rangle = \cos{ \left ( \frac{\alpha}{2} \right )}|L(t)\rangle+\sin{\left ( \frac{\alpha}{2} \right )}|R(t)\rangle,
\end{array}
\eeq
where the mixing angle $\alpha=\alpha(t)$ is given by $\tan \alpha = \delta (t)/\lambda$.

The bare basis states $\left \{ |L(t)\rangle,|R(t)\rangle \right\}$ are symmetrical and orthogonal moving left and right states. Initially when they
are close enough (and $\delta(0)>>\lambda$), the instantaneous eigenstates of $H$ are close to the symmetric ground state 
$|\psi^{-}_{0}(0)\rangle=\frac{1}{\sqrt{2}}(|L(0)\rangle+|R(0)\rangle)$
and the antisymmetric excited state 
$|\psi^{+}_{0}(0)\rangle=\frac{1}{\sqrt{2}}(|L(0)\rangle-|R(0)\rangle)$
of the single well.
At $t_f$ we should distinguish two extreme cases:
{\it{i)}} For $\delta(t_f)>>\lambda$ the final eigenstates of $H$ tend to $|\psi^{\mp}_{\lambda}(t_f)\rangle=\frac{1}{\sqrt{2}}(|L(t_f)\rangle\pm
|R(t_f)\rangle)$ which correspond to the symmetric and antisymmetric splitting states. 
{\it{ii)}} For  $\delta(t_f)<<\lambda$ the final eigenfunctions of
$H$ collapse and become right and left localized states: $|\psi^{-}_{\lambda}(t_f)\rangle=|L(t_f)\rangle$ and $|\psi^{+}_{\lambda}(t_f)\rangle=|R(t_f)\rangle$.
Since $\delta(t_f)$ is set as a small number to avoid tunnelling
in the final configuration, 
the transition from one to the other regime explains the sharp drop of $F_S$ at small $\lambda\approx\delta(t_f)$.

{\it{Dynamics of the two-mode model}.---}
%
%
%
%
%
\begin{figure}[t]
\includegraphics[width=6.cm]{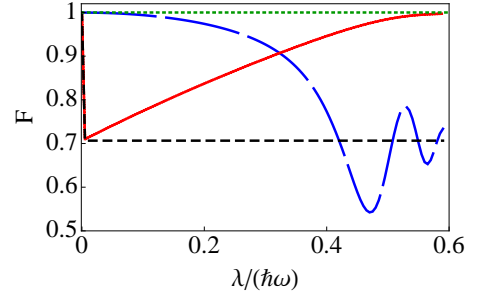}
\caption{(Color online) The same fidelities as in Fig. \ref{f2} versus the perturbation parameter $\lambda$ using the fast-forward approach for a Bose-Einstein
condensate. $F_D^{(0)}$: 
Blue long-dashed line; $F_D$: red line; $F_S$: black short-dashed line.
Parameter values: $t_f=320$ ms, $gN/(\hbar\omega a_{ho})=1.38$. The rest are the same as in Fig. \ref{f0}, ($a_{ho}=\sqrt{\hbar/(m\omega)}$).}
\label{f4}
\end{figure}
%
%
%
%
We define a moving-frame interaction-picture wave function  
$\psi^A=A^\dagger\psi^S$, where $A=\sum_{\beta=L,R} |\beta (t)\ra\la \beta(0)|$
and $\psi^S$ is the Schr\"odinger-picture wave function. 
In principle it obeys $i\hbar\dot{\psi}^A=(H_A -K_A) \psi^A$, with 
$H_A=A^\dagger H A$, and $K_A=i\hbar A^\dagger \dot{A}$ but, for real $\la x|R(t)\ra$ and $\la x|L(t)\ra$, the symmetry $\la x|R(t)\ra=\la -x|L(t)\ra$ makes
$K_A=0$.     
  
We may invert Eq. (\ref{eigenstates_tm}) to write the bare states in terms of the ground and excited states and energies,
and get $\delta (t)$ from the energy splitting in Eq. (\ref{eigenvalues_tm}). The
two-level model approximates the actual dynamics by first  
identifying $|\psi^{\pm}_0(t)\ra$ and $E^{\pm}_0(t)$ with the instantaneous ground and
excited states and energies of the unperturbed fast-forward Hamiltonian.
We combine them to compute the bare basis in coordinate representation. Then 
we compute the matrix elements $\la\beta'|H_{\lambda}|\beta\ra=H_{\lambda}^{\beta'\beta}$, ($H_{\lambda}=T+V_{\lambda}$) for $\beta\ne \beta'$. By comparison with Eq. (\ref{H_tm}) we
get $\delta(t)=-2H_{\lambda}^{RL}/\hbar=-2H_{\lambda}^{LR}/\hbar$. 
For the diagonal, $\beta=\beta'$, we may for consistency calculate  $\lambda'(t):=2(H_{\lambda}^{RR}-V_0)/\hbar=-2(H_{\lambda}^{LL}-V_0)/\hbar$, where $V_0=V_0(t)=
[E^-_\lambda(t)+E^+_\lambda(t)]/2$
is a shift to match the zero energy point between the fast-forward and the two-mode model. 
$\lambda'$ differs slightly from the constant $\lambda$ at short times.   
In our numerical calculations the results of substituting $\lambda$ by $\lambda'$ are hardly distinguishable in the final fidelities so the treatment with $\lambda$ is preferred 
for simplicity.   
Adjusting from the fast-forward approach the values of $\delta(t)$ 
we solve the dynamics in the moving frame for the two mode Hamiltonian. 
The initial state may be the ground state of the perturbed or unperturbed initial
potential, as for the fast-forward calculations 
the results can hardly be noticed in the figures. The
comparison with the exact results are excellent, see the symbols of Fig. \ref{f2},
so the two-level model provides a powerful interpretative and control tool. 
For further insight we shall perform further approximations.

{\it{Sudden and adiabatic approximations}.---}
The fidelities at low $\lambda$ may be understood with the sudden approximation \cite{Messiah}. Its validity
requires \cite{Messiah}
\beq
\label{sudden}
t_f \ll \hbar/\Delta \overline{H_A}, 
\eeq
where $\Delta \overline{H_A}=\sqrt{\langle\psi(0)|\overline{H_A}^2|\psi(0)\rangle-\langle\psi(0)|\overline{H_A}|\psi(0)\rangle^2}$. 
We take $|\psi(0)\rangle=|\psi_{0}^{-}(0)\rangle$, and $\overline{H_A}=\frac{1}{t_f}\int_{0}^{t_f}dt'H_A(t')$, where
the matrix elements of $H_A(t')$ in the bare basis $\{|\beta(0)\ra\}$
coincide with the matrix elements of $H$ in Eq. (\ref{H_tm}),
when the later are expressed in the  basis $\{|\beta(t')\ra\}$.
The condition for the sudden approximation to hold becomes 
\beq
\lambda \ll \frac{2}{t_f}.
\eeq
A vertical line marks $0.2/t_f$ in Fig. 2. 

The subsequent increase of $F_D$ for increasing $\lambda$ can be explained using the complementary adiabatic approximation. The adiabaticity condition is here \cite{Schiff, 3D}
\beq
|\langle \psi^-_\lambda(t)|\partial_t\psi^+_\lambda(t)\rangle| \ll \frac{1}{\hbar}|E^-_\lambda(t)-E^+_\lambda(t)|.
\eeq
%
Taking into account Eqs. (\ref{eigenvalues_tm}) and (\ref{eigenstates_tm}) 
it takes the form
\beq
\label{adiabaticity}
\left | \frac{\lambda \dot \delta(t)}{2(\lambda^2+\delta(t)^2)^{3/2}}
\right | \ll 1.
\eeq
%
%
%
Thus the structural and adiabaticity criteria delimit the small and large $\lambda$ values for which $F_D\approx 1$.
%
%
%
%
%
%
%
%
%
%
%
%
%
\begin{figure}[t]
\includegraphics[width=6.cm]{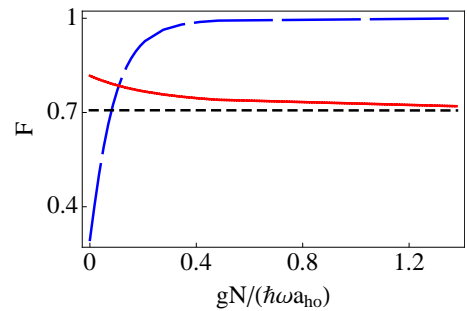}
\caption{(Color online) Fidelities for the Bose-Einstein condensate, 
same line codes as in Figs. 2 and 4.  
$\lambda/(\hbar \omega)=0.02$, 
$a=4$ $\mu$m, 
$t_f=320$ ms. 
}
\label{f5}
\end{figure} 

{\it{Bose Einstein condensates}.---} We use now the streamlined fast-forward
method to analyze the effects of the asymmetric perturbation in the splitting of a BEC. There is no analytical expression for the ground state of a BEC in a harmonic trap, so to mimic Eq. (\ref{ansatzrbueno}) 
we get first numerically the ground states $\chi_N(x)$ and $\chi_{\frac{N}{2}}(x)$ of a harmonic trap that holds a BEC with $N$ and $N/2$ particles and design the time evolution that connects these two states by interpolation as  
$f(x,t)=[1-{\cal{R}}(t)]\chi_N(x)+{\cal{R}}(t)\chi_{\frac{N}{2}}(x)$, where ${\cal{R}}(t)=3(t/t_f)^2-2(t/t_f)^3$. Finally $r(x,t)$ is 
constructed by displacing and summing these functions, 
\beq
r(x,t)=\frac{1}{z(t)}\bigg\{f[x-x_0(t),t]+f[x+x_0(t),t]\bigg\},
\eeq
where $z(t)$ is a normalization factor and $x_0(t)=a{\cal{R}}(t)$. Note that this form  reproduces Eq. (\ref{ansatzrbueno}) for $N=0$. We then get $V_{FF}$ from Eqs. (\ref{real}) and (\ref{imag}) and evolve in a perturbed potential $V_{\lambda}(t)=V_{FF}(t)+\lambda\theta(x)$ the initial ground state of the 
GP equation associated with $V_{\lambda}(0)$. 
The same fidelities as in the linear case may be computed and the results
are shown in Fig \ref{f4}. 
The structural fidelity (black short dashed line)
is not much different from the linear case, i.e., splitting by  adiabatic following is also very unstable for the condensate. Similarly, the dynamical fidelity $F_D$ (red solid line) drops abruptly at small $\lambda$ in a sudden regime to increase more slowly later on towards 
the adiabatic regime. There is however a very remarkable stabilization of $F_D^{(0)}\approx 1$
(blue long dashed line) with respect to the linear dynamics, as seen in Fig. \ref{f4} with respect to $\lambda$  (the scale for the $\lambda$-axis differs by an order of magnitude in Figs \ref{f2} and \ref{f4}) and Fig. \ref{f5} with respect to the non-linear coupling constant. In Fig. \ref{f5} we see that the relevant fidelity saturates to one. 
This means in summary that fast splitting via a designed fast-forward potential is significantly more robust versus the asymmetry for the condensate, as the non-linear term
compensates for the external potential asymmetry.   

{\it{Discussion}.---}
We have designed simple $Y$-shaped (position and time dependent) potential trap bifurcations to split matter waves rapidly without final excitation, avoiding the intrinsic instability 
of the adiabatic approach with respect to slight asymmetries. Incidentally, we also 
avoid or mitigate in this manner the decoherence effects that affect
slow adiabatic following.  
The bifurcation may be experimentally implemented by means of spatial light modulators \cite{modu}. A simpler approximate approach would involve the combination of Gaussian beams. 
Further standard manipulations may be combined with the proposed technique, in particular a differential phase among the two final parts may be imprinted by illuminating one of
them with a detuned laser.  

%
%
%
%
%
%
%
%
We are grateful to J. Martorell, A. Polls, B. Juli\'a-D\'{\i}az, D. Meschede,
and A. Aspect for fruitful discussions.  
We acknowledge funding by Projects No. GIU07/40 and No. FIS2009-12773-C02-01, 
and the UPV/EHU
under program UFI 11/55.  
E. T. acknowledges financial support from the Basque Government 
(Grants No. BFI08.151). X. C. thanks 
the National Natural Science Foundation of China (Grant No. 61176118).

\end{document}